\documentclass[fleqn,usenatbib]{mnras}

\usepackage{newtxtext,newtxmath}

\usepackage[T1]{fontenc}
\usepackage{ae,aecompl}

\usepackage{graphicx}	
\usepackage{amsmath}	

\title[The Sparkler Galaxy]{
Reconstructing the genesis of a globular cluster system at a look-back time of 9.1 Gyr with the {\it James Webb Space Telescope}} 

\author[D. A. Forbes \& A. J. Romanowsky]{
Duncan A. Forbes$^{1}$\thanks{E-mail: dforbes@swin.edu.au}
and Aaron J. Romanowsky$^{2,3}$
\\
$^{1}$Centre for Astrophysics \& Supercomputing, Swinburne University, Hawthorn, VIC 3122, Australia\\
$^{2}$Department of Physics \& Astronomy, San Jos\'e State University, One Washington Square, San Jose, CA 95192, USA\\
$^{3}$Department of Astronomy \& Astrophysics, University of California Santa Cruz, 1156 High Street, Santa Cruz, CA 95064, USA
}

\date{Accepted XXX. Received YYY; in original form ZZZ}

\pubyear{2019}

\begin{document}
\label{firstpage}
\pagerange{\pageref{firstpage}--\pageref{lastpage}}
\maketitle

\begin{abstract}

Using early-release data from the {\it James Webb Space Telescope}, Mowla et al. and Claeyssens et al. recently measured various properties for gravitationally lensed compact sources (`sparkles') around the `Sparkler' galaxy at a redshift of 1.378 (a look-back time of 9.1 Gyr). Here, we focus on the Mowla et al. as they were able to break the age-metallicity degeneracy and derive independent ages, metallicities and extinctions for each source.
They identified 5 metal-rich, old  
GC candidates (with formation ages up to $\sim$13 Gyr). 
 We examine the age--metallicity relation (AMR) for the GC candidates and other Sparkler compact sources. The Sparkler galaxy, which has a current estimated stellar mass of 10$^9$ M$_{\odot}$, is compared to the Large Magellanic Cloud (LMC), the disrupted dwarf galaxy Gaia--Enceladus and the Milky Way (MW). 
The Sparkler galaxy appears to have undergone very rapid chemical enrichment in the first few hundred Myr after formation, with its GC candidates similar to those of the MW's metal-rich subpopulation. 
We also compare the Sparkler to theoretical AMRs and formation ages from the E-MOSAICS simulation, finding the early formation age of its GCs to be in  some tension with these predictions for MW-like galaxies. 
The metallicity of the Sparkler's star forming regions are more akin to a galaxy of stellar mass $\ge$ 10$^{10.5}$ M$_{\odot}$, i.e. at the top end of the expected mass growth over 9.1 Gyr of cosmic time. 
We conclude that the Sparkler galaxy may represent a progenitor of a MW-like galaxy, even including the ongoing  accretion of a satellite galaxy.

\end{abstract}




\section{Introduction}

Globular clusters (GCs) are perhaps the oldest stellar systems in the Universe, 
formed only a few hundred Myr 
after the Big Bang (Gnedin 2004; Ricotti et al.\ 2016).  
It is not fully understood how they formed (see discussion in Forbes et al.\ 2018) but significant advances are being made, e.g.\  with the E-MOSAICS models for
the formation of GCs in a cosmological context using the EAGLE hydrodynamical simulations (Pfeffer et al.\ 2018).  
On the observational side, a large number of Milky Way (MW) GCs have age measurements from 
resolved colour--magnitude diagrams (e.g.\ Dotter et al.\ 2008; Marin-Franch et al.\ 2009). While finding very old ages, the systematic uncertainty is still large in such studies, i.e. around $\pm$ 1 Gyr.  

Combined with metallicities, one can probe the age--metallicity relation (AMR) of GCs in a galaxy and hence its chemical enrichment process. The shape of the AMR, or the level of enrichment at a given age, is correlated with the mass of the host galaxy (e.g.\ Horta et al.\ 2021). This fact has been used to deconstruct the accretion history of the MW, identifying the dwarfs that have been disrupted and the GCs deposited into the MW halo (Marin-Franch et al.\ 2009; Forbes \& Bridges 2010; Leaman et al.\ 2013; Kruijssen et al.\ 2019; Forbes 2020). The MW also contains a population of in-situ formed GCs which reveal significant chemical enrichment in a very short timeframe, i.e. enriching by $\sim$1.5 dex in metallicity within the first Gyr.
The origins of these GCs are unclear -- e.g.\ whether they were formed in early galaxy mergers or in a primordial turbulent disk.

An alternative approach to inferring the epoch of GC formation is to study GCs at a moderate redshift and hence significant look-back time. The GCs are then observed at an intermediate age which can be more accurately measured via spectroscopy or multi-filter photometry. Their chemical composition is set at birth and remains relatively unchanged over cosmic time. Thus the metallicity of GCs in distant galaxies can be directly compared to those locally. 
The challenge, of course, is that distant GCs are extremely faint. Magnification of GCs via strong gravitational lensing offers a potential break-through. Indeed, lensing of (relatively bright) proto-GCs has been studied using {\it HST} by Vanzella et al.\ (2017). 

Recently, Mowla et al.\ (2022; hereafter M22) studied the compact sources in and around the `Sparkler' galaxy using early-release {\it James Webb Space Telescope} ({\it JWST}) NIRCam imaging. 
M22 adopted a redshift for the Sparkler from its [O~II] emission line at $z = 1.378 \pm 0.001$. 
The galaxy is magnified by a factor of 5--100 by the foreground cluster SMACS0723 (at $z = 0.39$). Exploiting this magnification, and multiple lensed images, M22 measured fluxes in six broad bands for a total of 12 compact sources (`sparkles'). From spectral energy distribution (SED) fitting they estimated various physical properties including age, total metallicity, dust extinction and mass for each source. Compact sizes, red colours and inferred ancient formation epochs all indicated that half a dozen of the sources are excellent candidates for evolved GCs. 

The Sparkler thus provides a remarkable opportunity to study the properties of an emerging GC system in an adolescent universe ($\sim$1/3 its current age).
The host galaxy was estimated by M22 to have a de-magnified stellar mass of $M_\star \sim 10^9 \mathrm{M_\odot}$.  
Considering standard growth tracks for galaxies (Behroozi et al.\ 2019), the expected mass for the Sparkler at $z=0$ is $M_\star \sim 10^{10}  \mathrm{M}_\odot$ (with a variation of $\sim$0.5 dex),
and therefore the Sparkler is a possible progenitor for a MW-like mass 
($M_\star \sim 10^{10.5}  \mathrm{M}_\odot$)
galaxy.

Subsequent to M22, Claeyssens et al.\ (2022; hereafter C22) examined `clumps' in 18 lensed galaxies including the Sparkler. They measured photometry of the clumps and also conducted SED fitting. 
In this Letter we focus on the implications of the properties measured by M22 (ages, metallicities, masses, positions) for the compact
sources associated with the Sparkler galaxy.
However, we also briefly discuss the results of C22.
We adopt the same Cosmology as M22, i.e. $\Omega_M$ = 0.3, $\Omega_{\Lambda}$ = 0.7 and $H_0$ = 70 km s$^{-1}$ Mpc$^{-1}$, with the Big Bang occurring 13.7 Gyr ago, and $z = 1.378$ corresponding to a look-back time of 9.1 Gyr.

\section{The Ages and Metallicities of the Compact Sources in the Sparkler Galaxy}

From SED fitting of NIRCam photometry, M22 derived several physical parameters for the compact sources in and around the Sparkler galaxy. In particular, they measured non-parametric star formation histories (SFHs), ages, total metallicities, dust extinction and masses. Following Iyer et al.\ (2019), their Bayesian approach results in smooth SFHs and an age which is taken to be the time at which the SF peaks (denoted as t$_{\rm peak, 50}$ in their table 1). The latter has a resolution of around 0.5 Gyr.  They also performed
tests of injecting synthetic sources into the data and recovered ages with a systematic bias of 0.15 Gyr and a random scatter of 1 Gyr. Final ages and metallicities are quoted along with their 16th and 84th percentile confidence limits (which are consistent with their single stellar population fits).

M22 defined 5 compact sources to be GC candidates (i.e. objects 1,2,4,8,10). All 5 have red rest-frame colours of $u^*-r > 1.5$ indicative of a quenched stellar population.
They inferred an extinction of 0.23 $<$ A$_V$ $<$ 1.28 for the 5 objects.
They derived observed ages of $\ge$ 3.9 Gyr, or formation ages of $\ge$ 13 Gyr.  Total metallicities are in the range $-0.72 < \log Z/\mathrm{Z}_{\odot} < -0.11$ (similar to metal-rich GCs). 
Objects 11 and 12 are embedded in the main body of the galaxy and reveal [O~III] emission, indicating ongoing star formation (we exclude object 7 as its measured flux is about 1/10 that of 11 and 12 with correspondingly higher uncertainties). 
We include objects 5 and 6 which are listed as `extended' by M22 but appear to be compact sources with associated starlight (see discussion below). Following M22, we exclude objects 3 and 9 due to  contamination by nearby light. 



As a consistency check of the old ages and metal-richness of the GC candidates 
we use the online tunable SSP models of MILES\footnote{http://miles.iac.es/pages/webtools/tune-ssp-models.php}.
For a total metallicity of --0.4 (or --0.71) and age of 3.98 Gyr (close to the typical age found by M22) the models predict $u-r = 2.11$ (or $1.96$).
This is in good agreement with the typical colours measured by M22 for the GC candidates (see their figure 3) and  provides additional confidence in the stellar population parameters found by M22 from their SED fitting. 



In order to compare the total metallicities derived by M22 (denoted as log $Z_{50}/\mathrm{Z}_{\odot}$ in their table 1) with iron metallicities commonly used in the literature to study local GCs, we use the following conversion from 
Vazdekis et al.\ (2015): \\

\noindent
[Fe/H] = [$Z$/H] $-$ 0.75 [$\alpha$/Fe] \\

\noindent
In a review of MW GCs, Recio-Blanco (2018) 
compiled alpha-element ratios from the literature finding that GCs with ages $>$ 10 Gyr revealed $\alpha$-element ratios of 0.0 $<$ [$\alpha$/Fe] $<$ 0.5, with an average value of 0.3. Here we use [$\alpha$/Fe] = 0.3 to convert all of the Sparkler galaxy sources. 



\begin{figure}
	\includegraphics[width=0.8\columnwidth,angle=-90]{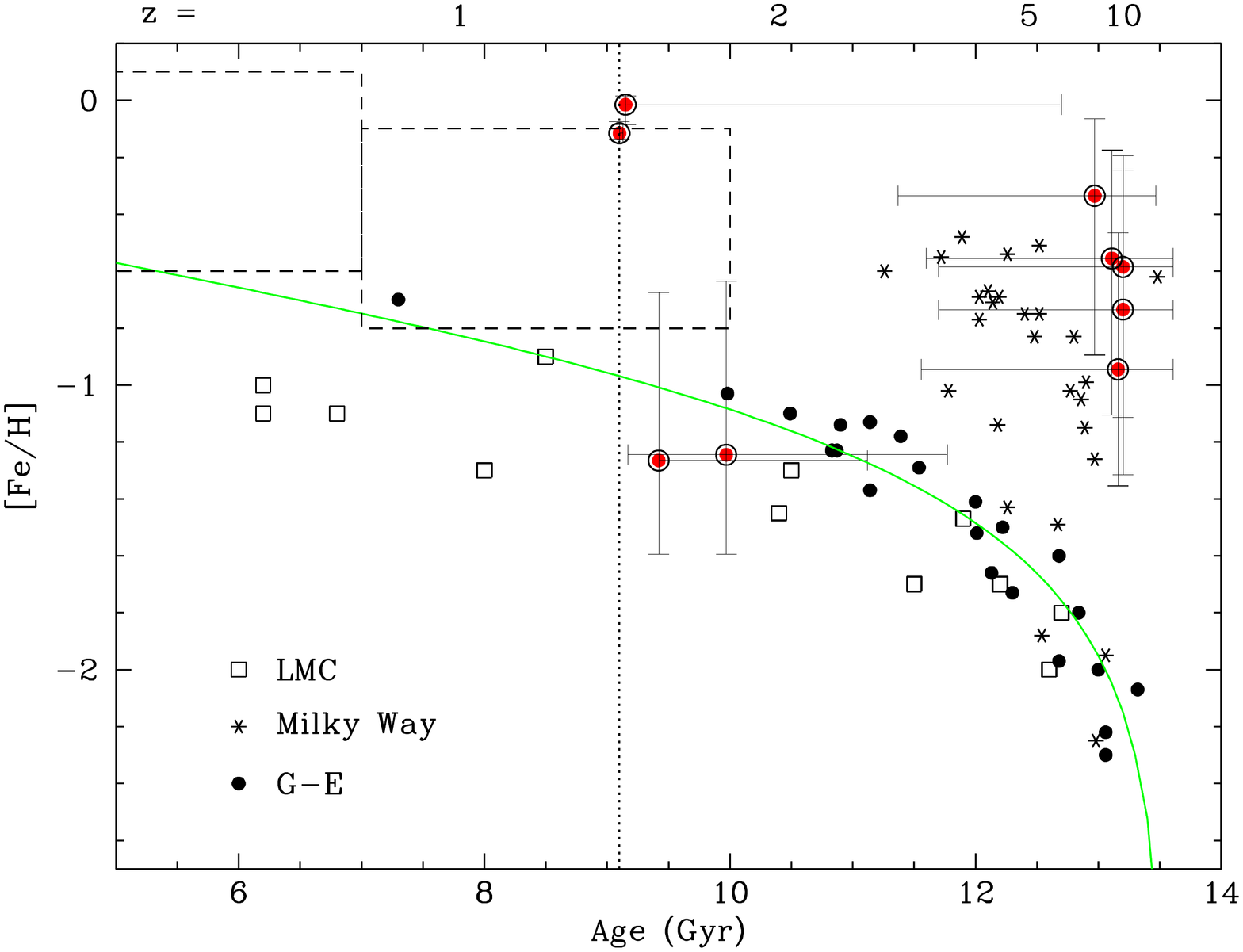}
    \caption{The Sparkler galaxy with a look-back time of 9.1 Gyr (dotted line) compared to observed age--metallicity relations (AMRs). Redshift is shown on the top axis. Sparkler globular cluster (GC) candidates (objects 1,2,4,8,10), star forming regions (objects 11,12) and extended sources (objects 5,6) from M22 are shown by large red circles. 
    GCs from the LMC (open squares) and Gaia--Enceladus (G--E, large black dots) are also shown. A leaky-box enrichment model for G--E GCs is represented by a solid green curve. The LMC and G-E galaxies have a similar stellar mass to the Sparkler (at $z = 1.378$). Asterisks indicate in-situ formed GCs of the MW and the dashed boxes the MW disk. The Sparkler galaxy has GCs similar to the metal-rich GCs of the MW but lacks its metal-poor GCs.  
    }
    \label{ge}
\end{figure}

In Fig.~\ref{ge} we show [Fe/H] versus age for the five GC candidates, two actively star forming regions and the two extended sources associated with the Sparkler using data from M22. Here age is the look-back time of 9.1 Gyr plus the derived t$_{\rm peak, 50}$ value. Error bars are the 16th and 84th percentile confidence limits from M22. The Sparkler old GCs have formation ages and [Fe/H] values similar to those of the metal-rich GC subpopulation of the MW. The two active star forming regions have ages consistent with the look-back time of the galaxy and near solar [Fe/H] metallicities. As the galaxy evolves over cosmic time we would expect it to grow in mass and to chemically enrich. Importantly, there is no expected change over cosmic time for the metallicity of its ancient GCs. The two extended sources have lower metallicities with formation ages of $\sim$ 10 Gyr. 

As noted above, the chemical enrichment of a galaxy as traced by its AMR is correlated with the galaxy stellar mass. 
The Sparkler currently has $M_\star \sim 10^9 \mathrm{M}_\odot$. For comparison 
we also show in Fig.~\ref{ge} the GCs of the LMC (with $M_\star = 2.7 \times 10^9$ M$_{\odot}$; Besla et al.\ 2015) and the AMR inferred from 28 GCs associated with the Gaia--Enceladus (G--E) dwarf galaxy (Forbes 2020). The G--E dwarf is perhaps the most massive dwarf galaxy accreted onto the MW, and is inferred to have a stellar mass of 
0.59--2.0 $\times$ 10$^9$ M$_{\odot}$ (and halo mass of 1--5 $\times$ 10$^{11}$ M$_{\odot}$) 
when it was accreted onto the Milky Way some 9--11 Gyr ago (Myeong et al.\ 2019; Forbes 2020). Thus it was accreted 
at a similar epoch to the current look-back time of the Sparkler, when it shared a similar mass.

The AMR of the old GCs and ongoing SF regions of the Sparkler lie well above those of the LMC and G--E to higher metallicities despite the oldest GCs in each galaxy forming at a similar epoch some 13 Gyr ago.  
Most of the Sparkler galaxy GCs have metallicities akin to those of the MW's (metal-rich) bulge GCs, 
while lacking 
the old metal-poor GCs seen in the MW, the LMC and the G--E galaxy. 
The Sparkler galaxy SF regions are similar to the metallicities seen in MW disk stars (Freeman \& Bland-Hawthorn 2002).
Bose \& Deason (2022) recently modelled the evolutionary paths of LMC analogues at $z = 2$, finding that proximity to a MW-like galaxy is the main factor determining whether the galaxy is accreted into its halo (like the G--E galaxy) or whether it survives largely intact until the present day (like the LMC).  

The sources 5 and 6 have measured compact sizes from C22 of $<$ 20 pc. M22 classified them as `extended' based on their `visual inspection'. Thus they appear to be unresolved sources with some associated diffuse starlight around them but located 
away from the galaxy main body. They have look-back ages of 
$\sim$10 Gyr and [Fe/H] $\sim$ --1.3, which places them in reasonable agreement with the AMR of the LMC and G--E galaxies.  We speculate that sources 5 and 6 are GCs formed some 10 Gyr ago in a low-mass satellite galaxy that is currently accreting onto the Sparkler galaxy. 

\begin{figure}
	\includegraphics[width=0.8\columnwidth,angle=-90]{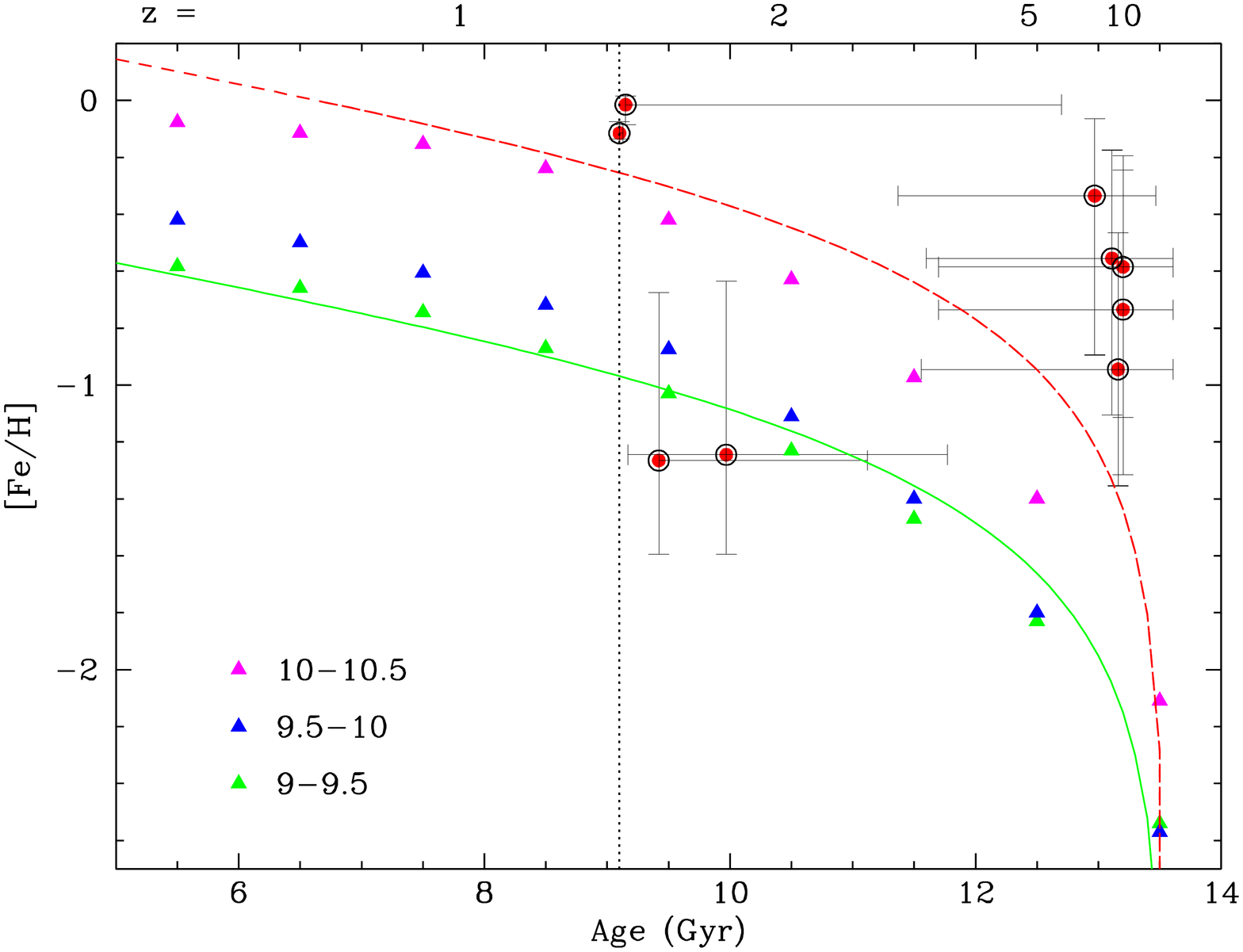}
    \caption{The Sparkler galaxy compared to model AMRs. Globular cluster candidates and two star forming regions in the Sparkler galaxy from M22 (red circles). 
    Coloured triangles show model $z=0$ AMRs from Horta et al.\ (2021) for stellar masses of $\log M_\star/\mathrm{M}_\odot$ = 9--9.5, 9.5--10 and 10--10.5 M$_{\odot}$. The MW-like models of Reina-Campos et al. (2019) predict metal-rich GCs to have a median age of  10.15 Gyr for --1 $<$ [Fe/H] $<$ --0.5. The green solid curve is the AMR of the G--E galaxy ($p$ = 0.27) and the dashed red line is an AMR with an effective yield of $p$ = 1.4. 
    The dotted vertical line is the look-back time of 9.1 Gyr.
    The Sparkler galaxy sources are most consistent with the highest mass AMR. 
    }
    \label{horta}
\end{figure}

Next we compare the Sparkler galaxy to model stellar AMRs from the E-MOSAICS simulation by Horta et al.\ (2021). These simulations are based on the EAGLE hydrodynamical simulations in a cosmological volume. We show their highest mass model AMRs  
in Fig.~\ref{horta} spanning stellar masses of 10$^{9-10.5}$ M$_{\odot}$ at $z = 0$ from 
figure 1 of Horta et al. (2021).
The lowest mass AMR shown corresponds very well to that of the G--E galaxy,
and to the two extended sources in the Sparkler galaxy -- consistent with these sources being recently accreted as part of a low-mass satellite galaxy.
%
The highest mass model AMR is reasonably consistent with the rest of the M22 Sparkler data, which supports the proposition that this high-redshift system will evolve into a galaxy of similar mass to the MW.  We note that the AMR for GCs tends to lie $\sim$0.2 dex above those of stellar AMRs such as produced by Horta et al. (J. Pfeffer, priv. comm.).


The AMR of a galaxy can be approximated by a leaky-box chemical enrichment model. The metallicity at a given time is given by:\\

\noindent
[Fe/H] = $-p$ ln ($\frac{t}{t_f}$),\\

\noindent
where $p$ is the effective yield of the system and $t_f$ is the look-back time when the system formed out of non-enriched gas. 
The AMR of the G--E galaxy shown in the figures has a best fitting yield of $p$ = 0.27$\pm$0.02 and $t_f$ = 13.55 Gyr (Forbes 2020). A similar yield and old formation age was also found recently by Limberg et al.\ (2022). 
In Fig.~\ref{horta} we show the AMR for the G--E galaxy along with a plausible AMR for the Sparkler galaxy, also with $t_f$ = 13.55 Gyr, but a 
yield of $p$ = 1.4, i.e. 5 times that of the G--E galaxy. Given the error bars on the M22 Sparkler values, the AMR is poorly constrained but it indicates that a yield much greater than that for a 10$^9$ M$_{\odot}$ stellar mass galaxy is required, as expected given the significant mass growth between z = 1.378 and z = 0.


As mentioned above, C22 have also measured photometry of the compact sources in the Sparkler galaxy and conducted SED fitting. 
They were unable to break the age-metallicity degeneracy and needed to hold the metallicity fixed to derive a best-fit age.
In versions 1 and 2 of their arXiv paper they quote derived ages for a fixed metallicity of [Z/H] = --0.04 (we assume Z$_{\odot}$ = 0.02) and variable extinction, and in version 3 they list in table F1 the best fitting age for three different metallicities 
For the GC candidates they additionally fix the extinction to zero (i.e. dust free). At the time of writing, their paper has yet to be accepted for publication. 
Qualitatively, C22 find similar age results to M22 (see their table B1 for their best-fit ages). In particular, they find sources within the main body of the galaxy to have the youngest ages, the embedded sources are slightly older and the 5 GC candidates have the oldest ages, up to $\sim$4 Gyr. However, from their  fixed metallicities they find the best fitting metallicity for the 5 GC candidates to be [Z/H] = --1.70, i.e. exclusively metal-poor compared to the findings of M22 for which all 5 are metal-rich. A visual inspection of figure F1 of C22 reveals that in some cases the best fit metallicity is not a particularly good fit to the observed photometry. 
Allowing for some extinction, and/or changes in other fitting parameters may improve the fit but a reanalysis of the C22 photometry is clearly beyond the scope of this paper. Nevertheless, the ages and metallicities of the 5 old GC candidates deserve further study.

\section{Discussion}

Here we have focused on the results from M22 but acknowledge the different results found by C22 for the same sources. While both studies found some old-aged GC candidates, the main difference is that M22 find the 5 old GC candidates to be significantly more metal-rich than claimed by C22 and hence rapid enrichment at early times. We find that the red u--r colours of the GC candidates are consistent with such metal-rich GCs. 
We note the degeneracy between metallicity and dust content in SED fitting, e.g. the more dust inferred, the lower the resulting metallicity (e.g.\ Jones \& Nuth 2011). So if the dust content has been underestimated by M22, 
then the true metallicities may be more metal-poor than shown in Figs. 1 and 2. 



Following M22, the compact sources in the Sparkler galaxy have relatively high metallicities. 
In particular, its oldest GC candidates have enriched very quickly (within a few hundred Myr of formation) to metallicities of [Fe/H] $\sim$ --0.5, i.e. similar to those associated with the MW's bulge. The AMR of the Sparkler suggests it is a galaxy with a high effective yield, and an average metallicity 
higher than might be expected from the simulated mass--metallicity relation of Ma et al.\ (2016). However, from the expected stellar mass growth over the next 9.1 Gyr from a current mass of 10$^9$ M$_{\odot}$ to a MW-like mass of 10$^{10.5}$ M$_{\odot}$, the Sparkler metallicities are similar to those predicted by the simulated AMR of Horta et al.\ (2021). 

The formation of GCs in a present day MW-like galaxy has been modelled in the E-MOSAICS simulation by Reina-Campos et al. (2019). In their 25 simulated MW galaxies, they found that the metal-rich (--1 $<$ [Fe/H] $<$ --0.5) GCs all have ages younger than 11.83 Gyr with a median age of 10.15.  Thus their simulations are in some tension with the Sparkler metal-rich GCs that have look-back ages of $\sim$ 13.5 Gyr and with quoted uncertainties of M22 that just reach their oldest predicted ages.  
Their predicted metal-poor GCs are around 2 Gyr older on average.





We next compare the GC mass distribution (GMCF) in the Sparkler to expectations for MW-like galaxies.  The `observed' GC masses are $M_\star \simeq$~1--5~$\times 10^8 \mathrm{M}_\odot$, which should be corrected by the lensing magnifications of $\sim$~5--100 (Mahler et al.\ 2022; Caminha et al.\ 2022).
The most massive metal-rich GCs in the MW have $M_\star \sim 10^6 \mathrm{M}_\odot$, so the highest range of these magnifications is required to avoid the Sparkler hosting implausibly massive GCs (as discussed by M22).
One implication here is that these GCs may be useful as a standard candle for constraining the lensing models, given the universality of the GC luminosity function. 

\begin{figure}
	\includegraphics[width=\columnwidth]{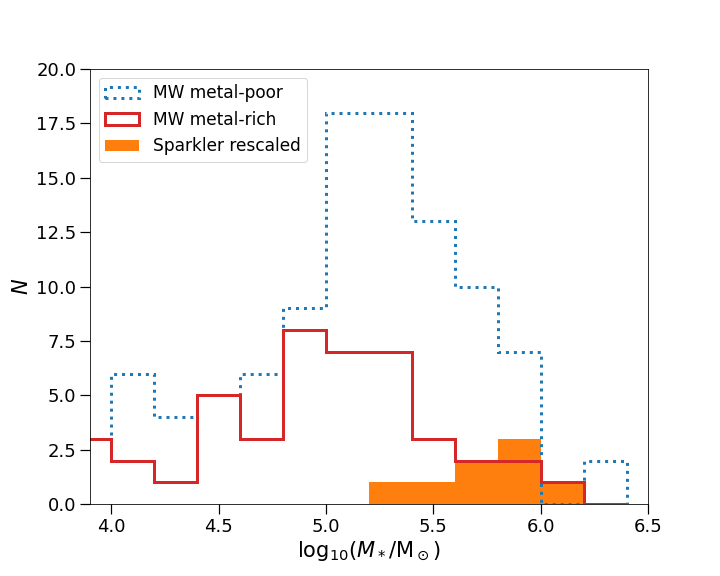}
    \caption{Globular cluster (GC) mass functions. The Sparkler galaxy including the GCs and extended sources (orange solid histogram, mostly metal-rich) is compared to the MW (metal-poor and metal-rich sub-populations shown as dotted blue and solid red histograms, respectively).  The Sparkler GC masses have been scaled down by a factor of 400, reflecting a combination of lensing magnification and mass-loss.
    }
    \label{gcmf}
\end{figure}

In more detail, Fig.~\ref{gcmf} compares the GCMFs of the Sparkler and the MW
(Harris 1996).  Here the MW GCs are divided into metal-rich and metal-poor at  [Fe/H]~$=-1.0$, and the Sparkler GC masses have been scaled
down by an arbitrary factor of 400 in order to match up with the MW at the high-mass end.  
This scaling could include a factor of a few to account for mass-loss between $z=1.378$ and $z=0$, and we note that the current sample of Sparkler
GCs is incomplete even for bright objects (e.g.\
there appears to be a good candidate nearby to object 8).
The re-scaled total mass of the observed Sparkler GCs is $M_\star \sim 6\times 10^6 \mathrm{M}_\odot$, which is $\sim$~10\% the mass of the MW GC system (Harris et al.\ 2013).
Deeper {\it JWST} imaging may reveal dozens more lower-mass old GCs in and around the Sparkler galaxy.


\begin{figure}
	\includegraphics[width=\columnwidth]{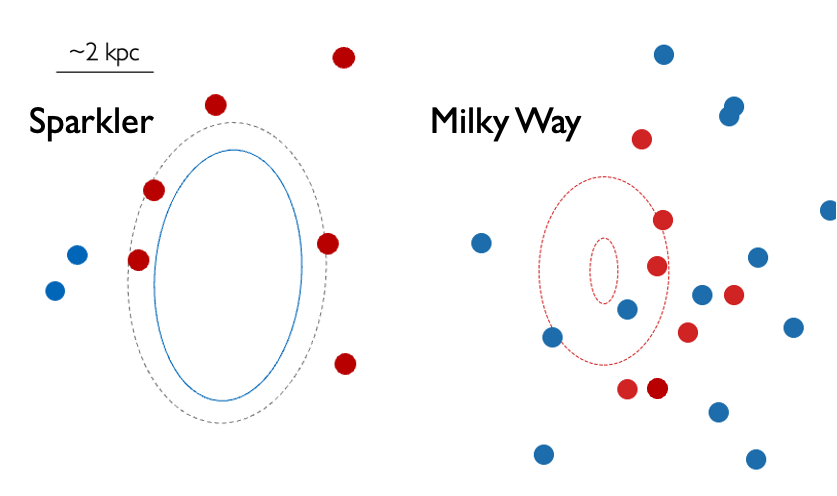}
    \caption{Schematic diagram of the Sparkler's spatial configuration compared to the MW, with a scale-bar shown for reference.
    Red and blue circles are metal-rich and metal-poor GCs, respectively, above a mass limit of $\sim 3\times10^5 M_\odot$.  Ellipses in the Sparkler outline the approximate boundaries of the brighter and fainter parts of the galaxy, using image 1 of M22 as the least distorted view of the morphology and GC spatial positions.
    The MW model is inclined to the line of sight by 54 degrees, and the ellipses outline the thick disk and bar (using scale length and height parameters from Bland-Hawthorn \& Gerhard 2016). The two galactic systems appear qualitatively similar, modulo dynamical friction effects expected on the Sparkler GCs by $z=0$, and the apparent absence of an old disk and bulge.
    }
    \label{spatial}
\end{figure}

A final observation concerns the spatial distribution of the GCs in the Sparkler.  There appears to be a striking decoupling between GCs and the host galaxy:  few if any GCs are seen embedded within the galactic disc, having projected galactocentric distances of $\sim$~2--4 kpc (after de-lensing and assuming a high magnification factor).
These distances are intriguingly similar to the 3D positions of the MW's metal-rich GCs, which are thought to have co-evolved with the galactic bulge and thick disk found on similar spatial scales
(see Figure~\ref{spatial}).
However, in the Sparkler it appears the GCs 
do not have 
a visibly associated galactic component
(unless the surface brightness of old stars is below the detection limit).
We conjecture that the GCs may have formed either through very early mergers or in a primordial disc that has been replaced by a younger disc with a shifted axis of angular momentum. In this case, extreme star formation conditions would have been required to form massive GCs with few accompanying field stars (e.g.\ Danieli et al.\ 2022).
We also note the recent high redshift simulations of Sameie et al. (2022) for dwarf galaxies which predicted GCs to form preferentially outside of the main body of the host galaxy.
An alternative is that GCs did form in the galaxy inner regions but later migrated to the outer regions, e.g.\ via mergers (Kruijssen 2015). 
We note also that the dynamical friction timescales for the GCs are $\sim$~5--10~Gyr,
so its GC system could evolve to become more compact and similar to the MW's by $z = 0$. 
If a lower magnification model were used, the implication would be that we are seeing massive
``ultracompact dwarfs'' rather than ordinary GCs, with even lower dynamical friction timescales, causing them to merge into the galaxy's central regions by the present day

\section{Conclusions}

Here we explore the ages and metallicities of several compact sources (`sparkles') in the Sparkler galaxy at $z = 1.378$ (look-back time = 9.1 Gyr). Such observations with {\it JWST} early-release data are made possible by strong lensing magnification factors of 5--100$\times$. Recently, 
independent studies by M22 and later C22 measured photometry of these compact sources and conducted SED fitting. These studies found similar results in terms of the ages of the compact sources with both agreeing that some GC candidates formed up to $\sim$13 Gyr ago. An unexplained difference is that M22 found that all 5 GC candidates are exclusively metal-rich while C22 favoured a metal-poor and dust-free solution in their best fit model.  
Focusing on the properties derived by M22 we compare the Sparkler sources to both observations in the local Universe and simulated AMRs. We find that the old GC candidates and young star forming regions are very metal-rich for their age indicating a very rapid chemical enrichment at early times. 

 If the Sparkler grows at the top end of the expected mass growth range over the next 9.1 Gyr, then its AMR is reasonably consistent with the simulations of Horta et al.\ (2021) using E-MOSAICS and the AMR inferred for the MW. Thus the Sparkler may be the progenitor of a MW-like mass galaxy. 
However, the E-MOSAICS simulation predictions for metal-rich GCs in MW-like galaxies are in some tension with the old ages inferred. 

We speculate that the Sparkler galaxy is currently accreting a low-mass satellite galaxy as evidenced by two sources of lower average metallicity beyond the main body of the galaxy. Deeper {\it JWST} observations may reveal a large number of lower mass old GCs.
More observations of the GC systems of galaxies spanning a range of intermediate redshifts and masses are needed to understand if the Sparkler is an atypical example or whether the details of galaxy chemical enrichment, mass growth and star cluster formation need some revision. 

\section*{Acknowledgements}

We thank D. Horta for supplying model data, and J. Gannon for his python help. We thank the ARC for financial support via DP200102574. 
We thank J. Gannon, J. Pfeffer and L. Buzzo for useful comments, and L. Mowla and K. Iyer for updated data for the Sparkler compact sources, and A. Claeyssens for some additional information regarding their analysis. 
We thank the referee for several useful comments.

\section*{Data Availability}

The data used in this project are taken from published works of Horta et al.\ (2021) and Mowla et al.\ (2022). However, reasonable requests for data can be made to DF.




\bibliographystyle{mnras}
\bibliography{MWGCs} 

\begin{thebibliography}{99}
\bibitem[Behroozi et al.(2019)]{2019MNRAS.488.3143B} Behroozi, P., Wechsler, R.~H., Hearin, A.~P., et al.\ 2019, \mnras, 488, 3143. 
\bibitem[Besla(2015)]{2015arXiv151103346B} Besla, G.\ 2015, arXiv:1511.03346
\bibitem[Bose \& Deason(2022)]{2022arXiv220905493B} Bose, S. \& Deason, A.~J.\ 2022, arXiv:2209.05493
\bibitem[Caminha et al.(2022)]{2022arXiv220707567C} Caminha, G.~B., Suyu, S.~H., Mercurio, A., et al.\ 2022, arXiv:2207.07567
\bibitem[Claeyssens et al.(2022)]{2022arXiv220810450C} Claeyssens, A., Adamo, A., Richard, J., et al.\ 2022, arXiv:2208.10450 (C22)
\bibitem[Dotter et al.(2008)]{2008AJ....136.1407D} Dotter, A., Sarajedini, A., \& Yang, S.-C.\ 2008, \aj, 136, 1407. 
\bibitem[Forbes, \& Bridges(2010)]{2010MNRAS.404.1203F} Forbes, D.~A., \& Bridges, T.\ 2010, \mnras, 404, 1203
\bibitem[Forbes et al.(2018)]{2018MNRAS.481.5592F} Forbes, D.~A., Read, J.~I., Gieles, M., et al.\ 2018, \mnras, 481, 5592
\bibitem[Forbes(2020)]{2020MNRAS.493..847F} Forbes, D.~A.\ 2020, \mnras, 493, 847. 
\bibitem[Freeman \& Bland-Hawthorn(2002)]{2002ARA&A..40..487F} Freeman, K. \& Bland-Hawthorn, J.\ 2002, \araa, 40, 487. 
\bibitem[Gnedin(2004)]{2004ASSL..301..215G} Gnedin, O.~Y.\ 2004, Astrophysics and Space Science Library, 301, 215. 
\bibitem[Horta et al.(2021)]{2021MNRAS.500.4768H} Horta, D., Hughes, M.~E., Pfeffer, J.~L., et al.\ 2021, \mnras, 500, 4768. 
\bibitem[Harris(1996)]{1996AJ....112.1487H} Harris, W.~E.\ 1996, \aj, 112, 1487. 
\bibitem[Harris et al.(2013)]{2013ApJ...772...82H} Harris, W.~E., Harris, G.~L.~H., \& Alessi, M.\ 2013, \apj, 772, 82. 
\bibitem[Iyer et al.(2019)]{2019ApJ...879..116I} Iyer, K.~G., Gawiser, E., Faber, S.~M., et al.\ 2019, \apj, 879, 116. 
\bibitem[Jones \& Nuth(2011)]{2011A&A...530A..44J} Jones, A.~P. \& Nuth, J.~A.\ 2011, \aap, 530, A44. 
\bibitem[Kruijssen(2015)]{2015MNRAS.454.1658K} Kruijssen, J.~M.~D.\ 2015, \mnras, 454, 1658. 
\bibitem[Kruijssen et al.(2019)]{2019MNRAS.486.3180K} Kruijssen, J.~M.~D., Pfeffer, J.~L., Reina-Campos, M., et al.\ 2019, \mnras, 486, 3180. 
\bibitem[Leaman et al.(2013)]{2013MNRAS.436..122L} Leaman, R., VandenBerg, D.~A., \& Mendel, J.~T.\ 2013a, \mnras, 436, 122
\bibitem[Limberg et al.(2022)]{2022ApJ...935..109L} Limberg, G., Souza, S.~O., P{\'e}rez-Villegas, A., et al.\ 2022, \apj, 935, 109. 
Mackey, D., Lewis, G.~F., Brewer, B.~J., et al.\ 2019, \nat, 574, 69
\bibitem[Ma et al.(2016)]{2016MNRAS.456.2140M} Ma, X., Hopkins, P.~F., Faucher-Gigu{\`e}re, C.-A., et al.\ 2016, \mnras, 456, 2140
\bibitem[Mar{\'\i}n-Franch et al.(2009)]{2009ApJ...694.1498M} Mar{\'\i}n-Franch, A., Aparicio, A., Piotto, G., et al.\ 2009, \apj, 694, 1498. 
\bibitem[Mahler et al.(2022)]{2022arXiv220707101M} Mahler, G., Jauzac, M., Richard, J., et al.\ 2022, arXiv:2207.07101
\bibitem[Mowla et al.(2022)]{2022ApJ...937L..35M} Mowla, L., Iyer, K.~G., Desprez, G., et al.\ 2022, \apjl, 937, L35. (M22) 
\bibitem[Myeong et al.(2019)]{2019MNRAS.488.1235M} Myeong, G.~C., Vasiliev, E., Iorio, G., et al.\ 2019, \mnras, 488, 1235
\bibitem[Pfeffer et al.(2018)]{2018MNRAS.475.4309P} Pfeffer, J., Kruijssen, J.~M.~D., Crain, R.~A., et al.\ 2018, \mnras, 475, 4309. 
\bibitem[Recio-Blanco(2018)]{2018A&A...620A.194R} Recio-Blanco, A.\ 2018, \aap, 620, A194
\bibitem[Reina-Campos et al.(2019)]{2019MNRAS.486.5838R} Reina-Campos, M., Kruijssen, J.~M.~D., Pfeffer, J.~L., et al.\ 2019, \mnras, 486, 5838. 
\bibitem[Ricotti et al.(2016)]{2016ApJ...831..204R} Ricotti, M., Parry, O.~H., \& Gnedin, N.~Y.\ 2016, \apj, 831, 204. 
\bibitem[Sameie et al.(2022)]{2022arXiv220400638S} Sameie, O., Boylan-Kolchin, M., Hopkins, P.~F., et al.\ 2022, arXiv:2204.00638
\bibitem[Vanzella et al.(2017)]{2017MNRAS.467.4304V} Vanzella, E., Calura, F., Meneghetti, M., et al.\ 2017, \mnras, 467, 4304. 
\bibitem[Vazdekis et al.(2015)]{2015MNRAS.449.1177V} Vazdekis, A., Coelho, P., Cassisi, S., et al.\ 2015, \mnras, 449, 1177. 


\end{thebibliography}





\bsp	
\label{lastpage}
\end{document}